# X-ray and neutron reflectometry study of copper surface reconstruction caused by implantation of high-energy oxygen ions

Yu. N. Khaydukov[1,2,3], O. Soltwedel[1,2], Yu. A. Marchenko[4], D. Yu. Khaidukova[5], A. Csik[6], T. Acartürk[1], U. Starke[1], T. Keller[1,2], A. G. Guglya[4], Kh. R. Kazdayev[7]

[1]Max Planck Institute for Solid State Research, Stuttgart, Germany
[2]Max Planck Society Outstation at FRM-II, D-85747 Garching, Germany
[3]Skobeltsyn Institute of Nuclear Physics, Moscow State University, Moscow, Russia
[4]Kharkiv Institute of Physics and Technology, Kharkiv, Ukraine
[5]Flerov Laboratory of Nuclear Reactions, JINR, Dubna, Russia
[6]Institute for Nuclear Research, Hungarian Academy of Sciences, Debrecen, Hungary
[7]LLP "Devir", Astana, Kazakhstan

**Abstract**
Combination of neutron and X-ray reflectometry was used to study the vertical structure of 100 nm-thin copper films with implanted oxygen ions of energy $E = [10 \div 30]$ keV and doses $D = [0.2 \div 5.4] \times 10^{16}$ cm$^{-2}$. The study shows that oxygen ion implantation with an energy of $E = 30$ keV leads to the formation of a 3 nm thick layer on the surface. Density and copper/oxygen stoichiometry of the observed surface layer are close to $Cu_2O$ oxide. We attribute the $Cu_2O$ oxide formation to highly mobilized copper atoms generated by stimulated ion implantation.

# Introduction

Interesting properties (low electrical resistivity, high thermal conductivity, high mechanical strength etc) promotes copper as widely used compound in many industrial and scientific applications. On the other side copper lacks of a self-passivating oxide films [1-10], as it is the case, for example, for aluminum or indium. Typically, as a result of oxidation a multilayered structure is formed on surface with composition $CuO/Cu_2O/Cu$ [2,4,7,9]. Several works reported about inhomogeneous surface of the oxide surface [7,9,11] with typical in-plane size of inhomogeneities of several microns.

Several methods improving copper passivation are suggested in the literature. Among them is the ion beam implantation, a method allowing to create a passivating layer of different composition on the surface of copper by bombarding a sample with different ions (C,N, D, O etc) with typical energies from eV to keV and implanted doses of $10^{15}$-$10^{17}$ cm$^{-2}$ [4,6,8,10].

For the study of surface oxidation several methods, like X-ray Photoemission spectroscopy (XPS), Secondary Ions Mass Spectrometry (SIMS), X-ray diffraction (XRD), spectroscopic ellipsometry, Auger electron spectroscopy (AES), Rutherford Backscattering Spectrometry (RBS) and Scanning Electron Microscopy (SEM) are widely used. XPS allows to define binding energies of surface elements and thus detects presence of copper and different oxides [1-3,6-8]. Since the crystallographic parameters of copper and its various oxides differs, XRD enables to investigate them in detail [2,8]. Methods based on transmission of X-ray beam through samples are typically not depth-selective. They allow detecting presence of a phase in sample but not the depth of its occurrence. Some depth-selectivity may be obtained by using of grazing incidence geometry. In this case the depth of the X-ray beam is adjusted by changing the grazing incidence angle of the incident beam. The grazing incidence diffraction [2,12] and absorption spectroscopy [9] are two examples. The disadvantage of this method is a low spatial resolution (>10nm). Depth selectivity can be obtained by analyzing of stepwise removed surface material using XPS, AES [3], SIMS or SNMS. These methods allow measuring depth profiles of elements with resolution of several nanometers along the surface normal, however in a destructive way. As a non-destructive method an example of RBS can be given. The method allows obtaining concentration profiles as a function of depth, expressed in relative units. To translate it into absolute units one needs additional information about the packing density of a layer.

In this work we used combination of neutron (NR) and X-ray reflectometry (XRR). These non-destructive methods utilize dependence of the scattered intensity on the depth variation of the scattering length density (SLD) of the investigated structure. Since neutrons scatter on nucleus and X-rays on electron shells, the SLD profiles of neutrons and X-rays are significantly different. Typical reflectivity curve consists of a plateau of total external reflection at $Q < Q_{cr}$ and Kiessig oscillations with period $dQ$. Latter depends on the thickness of a layer $d$ as $dQ \approx 2\pi/d$. Reflectivity curves from multilayered structures contain different oscillations. The critical edge is associated with the scattering length density $\rho$ of a layer as $Q_{cr} = (16\pi\rho)^{1/2}$. Thus comprehensive analysis of XRR and NR allows experimentally to restore both concentration profiles (as in case of using SIMS, SNMS, AES) and, what is unique, the depth profile of density with spatial resolution of one nanometer.

## Sample preparation and characterization

Samples were prepared by vapor deposition in Kharkiv Institute of Physics and Technology on Si(111) substrates with thickness 0.5mm. Residual pressure during preparation was $4\times10^{-6}$ mbar. One sample of nominal structure Cu(100nm)/Si was prepared as virgin, untreated sample (*pCu*). Others were treated by oxygen ion beam with energies $E = [10\div30]$ keV and doses $D = [0.2\div5.4] \times 10^{16}$ cm$^{-2}$.

Structural properties of the samples were tested by X-ray and neutron reflectometry on combined neutron/X-ray reflectometer NREX located at the research reactor FRM II (Munich, Germany). For measurements of neutron and X-ray reflectivities neutron beam with $\lambda = 4.3$ Å and X-ray beam with $\lambda = 1.54$ Å was used. The depth profiles of elemental concentrations of part of the samples were measured by SIMS and SNMS.

Fig. 1a shows the XRR and NR reflectivity curves measured on the untreated sample pCu. Both X-ray and neutron reflectivity curves are characterized by the total reflection plateau with critical edge $Q_{cr}$ and Kiessig oscillations with period $dQ \approx 0.7$ nm$^{-1}$ caused by interference on the layer of Cu layer with thickness $d \approx 2\pi/dQ \approx 100$nm. Both XRR and NR experimental curves were fitted using a simple model of a single homogeneous layer (Fig. 1c,d). Within this model the SLD profile was parameterized using thickness of the Cu layer $d$, X-ray and neutron SLDs $\rho_X$ and $\rho_n$ and root-mean-square roughness on the interfaces air/Cu and Cu/Si $\sigma_1$ and $\sigma_2$ correspondingly. The SLDs corresponding to the best-fits, $\rho_n = 6.6\times10^{-4}$ nm$^{-2}$ and $\rho_X = 62.7\times10^{-4}$ nm$^{-2}$ are within 2% in agreement with literature values (assuming bulk copper mass density: 8.96 g·cm$^{-3}$), proving thus high quality of initial copper layer. The r.m.s. roughness of air/Cu and Cu/Si interface were found to be around 2nm. Additional SIMS measurements proved absence of oxygen in the copper layer except interface regions (see inset in Fig. 1c).

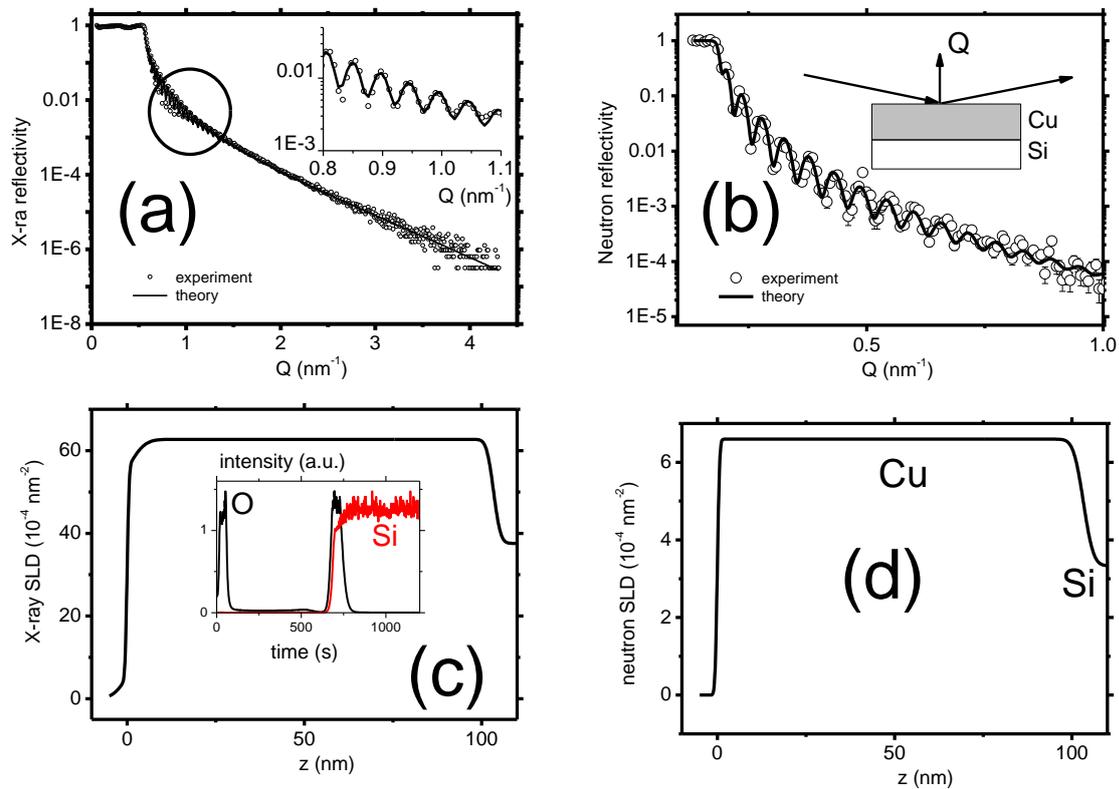

Fig. 1. Experimental (dots) and model (solid lines) X-ray (a) and neutron (b) reflectivity curves measured on the virgin sample pCu. Inset in (a) shows the zoomed reflectivity around $Q = 1$ nm$^{-1}$ where Kiessig oscillations with period 0.7 nm$^{-1}$ are better seen. Inset in (b) shows scheme of reflectometry experiment. (c). The X-ray SLD depth profile corresponding to the fit shown in Fig 1a. Inset shows the SIMS spectra for oxygen (black) and silicon (red) atoms. The latter is shown to determine position of the substrate. (d) Extracted neutron SLD profile corresponding to the fit shown in Fig 1b .

The XRR and NR curves for the treated samples are shown in Fig. 2. The main difference clearly seen on the XRR curves for the samples S1 and S2 is the presence of another Kiessig oscillations with period of order $dQ_2 \approx 2nm$. To describe this "hump" an additional layer with thickness $\Delta d$ and SLDs for X-rays and neutrons $\Delta \rho_X$ and $\Delta \rho_n$ was introduced in the model. Fit procedure was as follows. First the XRR curve was fitted varying thicknesses and SLDs of both layers. Next, the NR curve was fitted by varying only SLDs of both layers keeping the thicknesses from the fit of the XRR. The finally obtained SLD profiles for S1 are shown in insets to Fig. 2. Parameters for all structures are summarized in Table 1. We also tried to describe the hump on XRR curves by introducing a layer on the substrate side, but this model does not reproduce experiment adequately.

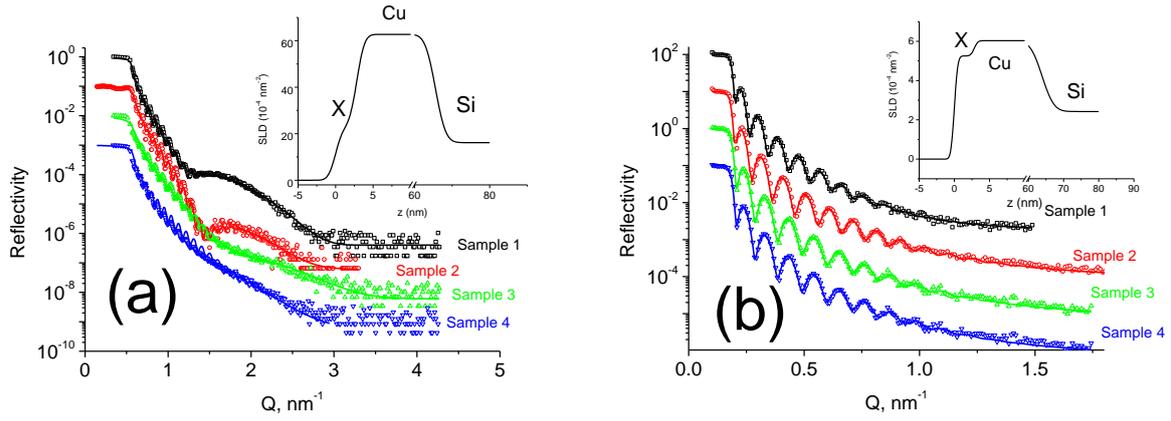

Fig. 2. Experimental (dots) X-ray (a) and neutron (b) reflectivities after ion implantation as described in Table 1. The scattering length density profiles corresponding to the model X-ray and neutron reflectivities are depicted in insets.

Knowing $\Delta\rho_X$ and $\Delta\rho_n$ and assuming that this layer consists of Cu and O only, enables to extract the copper concentration ($C_{Cu}$) and atomic density of the layer ($N$). Then, obviously, for a layer consisting of Cu and O the SLDs can be written as

$$\Delta\rho_X = N[C_{Cu}\cdot b_{Cu,X}+(1-C_{Cu,X})\cdot b_{O,X}] \qquad (1)$$

$$\Delta\rho_n = N[C_{Cu}\cdot b_{Cu,n}+(1-C_{Cu,n})\cdot b_{O,n}] \qquad (2)$$

Here $b_{Cu,X} = 81.78$ fm, $b_{O,X} = 22.56$ fm are scattering lengths of X-rays from copper and oxygen, $b_{Cu,n} = 7.72$ fm, $b_{O,n} = 5.80$ fm are scattering lengths for neutrons. By solving equations (1) and (2) the density and concentration of copper and oxygen atoms can be calculated (Table 1). As it follows from Table 1 the concentration of copper in the surface layer of samples S1 and S2 corresponds to stoichiometry $Cu_2O$. The experimentally obtained density for samples S1 and S2 is of the order 6.0 g/cm$^3$, close to the bulk value for $Cu_2O$ [9]. In Ref. 12 density of $Cu_2O$ layer on Cu(35nm) film of order 6.1 g/cm$^3$ is derived from XRR curve fitting. However, as it is shown above, using of only XRR is not enough to calculate experimental density of oxide layer. For samples 3 and 4 the calculation gives inadequate parameters $C_{Cu}$ and $N$. We relate this with small amplitude of the Kiessig oscillations on the XRR curve.

Table 1. Parameters of the samples and XRR+NR fit results.

| ID | E, keV | Dose cm$^{-2}$ | $\Delta d$, nm | $\Delta\rho_X$ 10$^{-4}$ nm$^{-2}$ | $\Delta\rho_n$ 10$^{-4}$ nm$^{-2}$ | $C_{Cu}$, % | N A$^{-3}$ |
|---|---|---|---|---|---|---|---|
| S1 | 30 | 2x10$^{15}$ | 3.0 | 46.67 | 5.25 | 69 | 0.0739 |
| S2 | 30 | 8x10$^{15}$ | 2.7 | 49.50 | 5.58 | 68 | 0.0785 |
| S3 | 10 | 2x10$^{16}$ | 2.4 | 50.20 | 0.32 | -370 | -0.0256 |
| S4 | 15 | 2x10$^{16}$ | 1.9 | 16.97 | 6.55 | -14 | 0.1182 |

The formation of oxide $Cu_2O$ instead of CuO on the surface of copper after ion bombardment was shown in Ref. [6]. In this study high-energy 200keV ions of copper were used. These ions produce a large number of point defects (around 3 displacements per atom, dpa), which increase the diffusion of copper atoms to the surface. The increased copper-ion mobility creates an excess of Cu atoms on the surface. Energy transfer simulations made by SRIM program [13] for oxygen ions also provide the presence of point defects with high density (Fig. 3). The difference lies in the fact that ions with energy of 30keV create point defects throughout the whole layer of copper, while the 10 keV and 15keV only near the surface. Thus, samples 1 and 2 possess large number of point defects throughout the whole layer, providing transport of copper ions to the surface. This leads to the formation of $Cu_2O$ oxide.

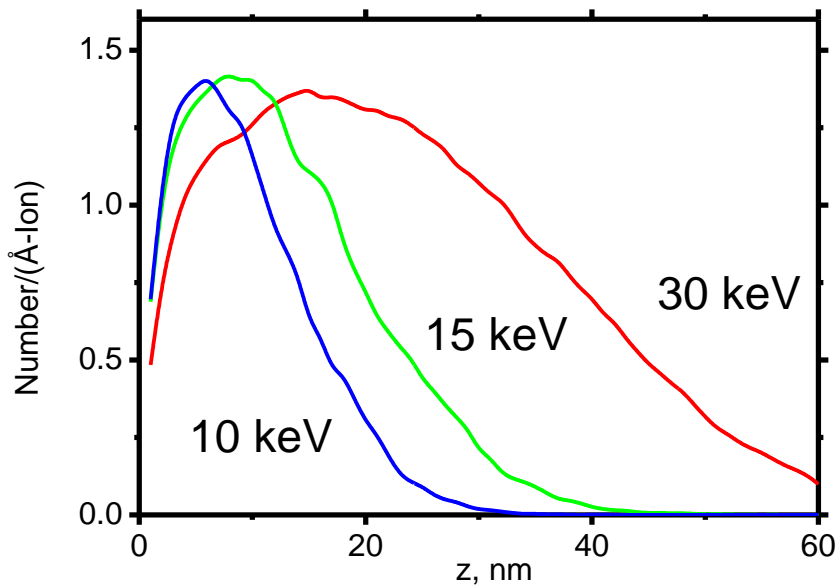

Fig. 3. The depth profile of the displaced atoms per ion per Anstroem calculated by SRIM program.

In conclusion, combination of neutron and X-ray reflectometry was used to study structure of copper films with implanted oxygen ions of energy $E = [10 \div 30]$ keV and doses $D = [0.2 \div 5.4] \times 10^{16}$ cm$^{-2}$. The study allowed to define that implantation of ions with the energy $E = 30$ keV leads to the formation of 3nm thick layer with density and copper/oxygen stoichiometry close to $Cu_2O$ oxide. Formation of $Cu_2O$ oxide can be related to implantation stimulated high mobility of copper atoms.

### Acknowledgements

We are grateful to Dr. Harm Wulff (Greifswald University, Germany) for fruitful discussions of the results. Assistance of Dr. Anatoly Senyshin (Heinz Maier-Leibnitz Zentrum, Germany) in preparation of the experiment is also acknowledged. This work is based upon experiments performed at the NREX instrument operated by Max-Planck Society at the Heinz Maier-Leibnitz Zentrum (MLZ), Garching, Germany. The SNMS depth profile measurement was carried out as part of the TÁMOP-4.2.2.A-11/1/KONV-2012-0019 project in the framework of the New Széchenyi Plan. The realization of this project is supported by the European Union, co-financed by the European Social Fund. This work is dedicated to the memory of Dr. Khamza Kazdayev, talented scientist and good man, who initiated this study, but passed away in 2013.